\documentstyle[aps]{revtex}

\begin{document}
\author{A.I.Buzdin and H.Kachkachi}
\address{Centre de Physique Th\'{e}orique et de Mod\'{e}lisation, \\
Universit\'{e} Bordeaux I, CNRS-URA 1537\\
F-33174 Gradignan Cedex, France.}
\title{Generalized Ginzburg-Landau theory for non-uniform FFLO superconductors }
\maketitle

\begin{abstract}
We derive a generalized Ginzburg-Landau (GL) functional near the tricritical
point in the $(T,H)$-phase diagram for the Fulde-Ferrell-Larkin-Ovchinnikov
(FFLO) superconducting state, in $1,2$, and $3$ dimensions. We find that the
transition from the normal to the FFLO state is of second order in $1$ and $%
2 $ dimensions, and the order parameter with one-coordinate sine modulation
corresponds to the lowest energy near the transition line. We also compute
the jump of the specific heat and describe in the one-dimensional case the
transformation of the sine modulation into the soliton-lattice state as the
magnetic field decreases. In $3$ dimensions however, we find that the
transition into an FFLO state is of first order, and it is impossible to
obtain an analytic expression for the critical temperature. In this case the
generalized GL functional proposed here provides a suitable basis for a
numerical study of the properties of the FFLO state, and in particular for
computing the critical temperature, and for describing the transition into a
uniform state.



\vspace*{0.6cm}

Corresponding author: kachkach@bortibm4.in2p3.fr
\end{abstract}

\section{Introduction}

As it was shown long time ago by Larkin and Ovchinnikov\cite{LO} and by
Fulde and Ferrell \cite{FF}, at low temperatures and when the magnetic field
is acting on the spin of electrons only, a transition from normal (N) to
modulated superconducting state (FFLO state) must occur. Due to this
non-uniform superconducting state formation the paramagnetic limit at $T=0$
becomes larger than the usual Chandrasekhar-Clogston limit\cite{SST}, $%
H_{p}(0)=\frac{\Delta _{0}}{\mu _{B}\sqrt{2}}$, where $\Delta
_{0}=1.76\,T_{c}$ is the superconducting gap at $T=0$. The $(T,H)$ phase
diagram for $3D$ superconductors was obtained by Saint-James and Sarma\cite
{SST} assuming that the transition $N\rightarrow FFLO$ is of second order.
It happens that the FFLO state only appears at $T<T^{*}\simeq 0.56\,T_{c}$%
\cite{SST} and that the temperature-dependence of the critical field $%
H^{FFLO}(T)$ is strongly influenced by the dimensionality of the system.
Indeed, for instance in the $3D$ case $H_{3D}^{FFLO}(0)=0.755\,\frac{\Delta
_{0}}{\mu _{B}}$, and $H_{2D}^{FFLO}(0)=\,\frac{\Delta _{0}}{\mu _{B}}$ for $%
2D$ superconductors \cite{Bulaev}, while this field diverges in the
one-dimensional case \cite{BT}.

Up to now there is no conclusive experimental evidence for the FFLO state
formation, except perhaps for what concerns $UBe_{13}$\cite{Thomas}.
However, for this heavy-fermion superconductor the applicability of the
standard theory of superconductivity is not evident. The main reason for the
difficulties in observing such state experimentally resides in the fact that
the orbital effect is usually more important than the paramagnetic one, and
that the actual critical field is mainly determined by the orbital effect.
However, for heavy-fermion superconductors and low-dimensional
superconductors (when the field is applied parallel to the planes or chains)
the orbital effect can be suppressed, and thence we deal with
paramagnetically limited critical field.

The problem of exact structure of the FFLO state is not solved yet even in
the framework of the model of pure paramagnetic limit, except for the $1D$
case where the superconducting order parameter in the FFLO sate is described
by the Jacobi elliptic function\cite{BT}, \cite{BP}, \cite{MN}. For this
reason in this paper we concentrate on the description of the FFLO phase in
the vicinity of the tricritical point $(T^{*},H_{p}(T^{*}))$ where the
characteristic wavevectors of the FFLO state are small compared with the
inverse superconducting coherence length $\xi _{0}^{-1}$. In this case, the
appearance of the non-uniform state is related with a change of the sign of
the coefficient $\beta $ at the gradient term $\beta (\nabla \psi )^{2}$ in
the free energy. In the standard Ginzburg-Landau (GL) theory the coefficient 
$\beta $ is positive, but it happens to be a function of the external field
acting on the electron spins (paramagnetic effect) and vanishes at the point 
$(T^{*},H_{p}(T^{*}))$, and then becomes negative for $T<T^{*}$. A negative $%
\beta $ means that the modulated state corresponds to lower energy as
compared with the uniform one. So, in order to obtain the modulated vector
one needs to include a term with second derivative in the GL functional.
Moreover, in BCS theory the vanishing of the gradient term at $%
(T^{*},H_{p}(T^{*}))$ is simultaneously accompanied with that of the
coefficient $\gamma $ of the fourth-order term $\gamma \psi ^{4}$\cite{SST}.
Due to this particular property, one needs to add higher-order terms such
as, $\psi ^{6}$ and $(\nabla \psi )^{2}\psi ^{2}$. Note that some particular
cases of generalized GL functional were considered in \cite{BP},\cite{BK}.

In this article we derive a generalized GL functional for $1,\;2$ and $3D$
superconductors which provides a thorough description of the FFLO state near
the tricritical point. Here, we find that in one and two dimensions the
transition from the normal to the FFLO superconducting state is of second
order while it is of first order in the three-dimensional case. In any case
the state with simple exponential modulation of the order parameter, i.e. $%
\psi \sim \left| \psi _{0}\right| \;e^{iq\cdot r}$, considered in ref.\cite
{TI}, does not yield the minimum energy and always corresponds to a
second-order phase transition. In the $1D$ and $2D$ cases it is most
favourable for the FFLO state to appear with a simple sine structure, i.e. $%
\psi \sim \psi _{0}\sin (q\cdot r)$. For $2D$ superconductors such structure
has lower energy when compared with the ''$2D$ lattice'', $\psi \sim \psi
_{0}\;\left( \sin (q\cdot x)+\sin (q\cdot y)\right) $, a result that
justifies the choice made in ref.\cite{Burkh} of the FFLO structure for
layered superconductors with an order parameter depending only on one
coordinate. For $3D$ superconductors, however, the situation turns out to be
more complicated. In this case we find that the transition is of first order
for the sine state with one, two, and three-dimensional modulation. Our
approach yields a good basis for a numerical study of the structure and
properties of the FFLO state.

\section{Generalized Ginzburg-Landau functional}

In the paramagnetic limit the Hamiltonian of the system can be written in
the mean-field approximation, in a $d$-dimensional space as follows 
\begin{equation}
H=\int d^{d}r\left\{ 
\begin{array}{c}
\mathop{\displaystyle \sum }%
\limits_{\sigma }\left[ \Phi _{\sigma }^{+}(r)\frac{\nabla ^{2}}{2m}\Phi
_{\sigma }(r)+\sigma {\cal H}\;\Phi _{\sigma }^{+}(r)\Phi _{\sigma
}(r)\right] + \\ 
\left( \psi (r)\cdot \Phi _{1}^{+}(r)\Phi _{-1}^{+}(r)+h.c.\right)
\end{array}
\right\}  \label{L0}
\end{equation}
where $\sigma =+1(-1)$ if the electron spin is parallel (anti-parallel) to
the magnetic field $H$ and ${\cal H}=\mu _{B}H$. The superconducting order
parameter is $\psi (r)=\lambda \left\langle \Phi _{1}(r)\Phi
_{-1}(r)\right\rangle $, where $\lambda $ is the electron-phonon coupling
constant.

Near the transition line the order parameter $\psi $ is small, thus using
the Gorkov procedure\cite{AGD} for deriving the Ginzburg-Landau functional,
we write 
\begin{eqnarray*}
F &=&\int d^{d}r\frac{\left| \psi \right| ^{2}}{\left| \lambda \right| }-%
\frac{1}{\beta }\sum_{\nu }\int d^{d}r_{1,2}\;G^{-}(-\omega _{\nu
},r_{1}-r_{2})G^{+}(\omega _{\nu },r_{2}-r_{1})\psi (r_{1})\psi ^{+}(r_{2})
\\
&&\ \ +\frac{1}{2\beta }\sum_{\nu }\int d^{d}r_{1,2,3,4}\;G_{\omega
}^{+}(r_{4}-r_{1})G_{\omega }^{-}(r_{4}-r_{2})G_{\omega
}^{-}(r_{3}-r_{1})G_{\omega }^{+}(r_{3}-r_{2})\ \times \\
&&\ \ \psi (r_{1})\psi (r_{2})\psi ^{+}(r_{3})\psi ^{+}(r_{4})+\cdots
\end{eqnarray*}

where the Green functions here are defined by 
\begin{eqnarray*}
G^{+}(\omega _{\nu },p) &=&\frac{1}{i\omega _{\nu }-\xi _{p}+{\cal H}},\quad
\;G^{-}(\omega _{\nu },p)=\frac{1}{i\omega _{\nu }-\xi _{p}-{\cal H}} \\
\xi _{p} &=&\frac{p^{2}}{2m}-\varepsilon _{F}
\end{eqnarray*}

Being close to the tricritical point of the $(T,H)-$phase diagram, the
spatial modulation of the order parameter is small so that we can expand $%
\psi (r_{i})$ around $r_{1}$ in Taylor series 
\begin{equation}
\psi (r_{i})\simeq \psi (r_{1})+\left( (\overrightarrow{r}_{i}-%
\overrightarrow{r}_{1})\cdot \overrightarrow{\nabla }\right) \psi +\frac{1}{%
2!}\left( (\overrightarrow{r}_{i}-\overrightarrow{r}_{1})\cdot 
\overrightarrow{\nabla }\right) ^{2}\psi +O(3)  \label{L3}
\end{equation}

After long but straightforward calculations, the free energy can be
rewritten as\footnote{%
We have also added the sixth-order term because the fouth order one turns
out to be small near the tricritical point.} 
\begin{eqnarray}
F &=&\alpha \left| \psi \right| ^{2}+\beta \left| \partial \psi \right|
^{2}+\gamma \left| \psi \right| ^{4}+\delta \left| \partial ^{2}\psi \right|
^{2}+\mu \left| \psi \right| ^{2}\left| \partial \psi \right| ^{2}
\label{GLF} \\
&&+\eta \left[ (\psi ^{+})^{2}(\partial \psi )^{2}+\psi ^{2}(\partial \psi
^{+})^{2}\right] +\nu \left| \psi \right| ^{6}  \nonumber
\end{eqnarray}
with the coefficients 
\[
\alpha =-\pi N(0)\cdot (K_{1}-K_{1}^{0}),\quad \gamma =\frac{\pi N(0)K_{3}}{4%
},\quad \nu =-\frac{\pi N(0)K_{5}}{8} 
\]
in all dimensions, and in $1D$ 
\[
\beta =\frac{\pi N(0)V_{F}^{2}K_{3}}{4},\quad \delta =-\frac{\pi
N(0)V_{F}^{4}K_{5}}{16},\quad \mu =8\eta =-\frac{\pi N(0)V_{F}^{2}K_{5}}{2} 
\]
in $2D$%
\[
\beta =\frac{\pi N(0)V_{F}^{2}K_{3}}{8};\quad \delta =-\ \frac{3}{8}\frac{%
\pi N(0)V_{F}^{4}K_{5}}{16},\quad \mu =8\eta =-\frac{\pi N(0)V_{F}^{2}K_{5}}{%
4}, 
\]
and in $3D$ 
\[
\beta =\frac{\pi N(0)V_{F}^{2}K_{3}}{12};\quad \delta =-\ \frac{\pi
N(0)V_{F}^{4}K_{5}}{80},\quad \mu =8\eta =-\frac{\pi N(0)V_{F}^{2}K_{5}}{6}%
\quad 
\]
where $V_{F}$ is the Fermi velocity, $N(0)$ the electron density of states,
and 
\begin{eqnarray*}
K_{n} &=&2T\cdot 
\mathop{\rm Re}%
\left[ \sum_{\nu =0}^{\infty }\frac{1}{(\omega _{\nu }-i{\cal H})^{n}}%
\right] =\left( 2T\right) ^{n-1}\frac{1}{\pi ^{n}}%
\mathop{\rm Re}%
\left[ \sum_{\nu =0}^{\infty }\frac{1}{(\nu +z)^{n}}\right] ,\;n\geq 1 \\
&& \\
\omega _{\nu } &=&\pi (2\nu +1)T,\quad z=\frac{1}{2}-i\frac{{\cal H}}{2\pi T}
\end{eqnarray*}
or in terms of the {\it psi} function $\Psi (x)=\frac{d}{dx}\ln \Gamma (x) $%
, $K_{3}$ and $K_{5}$ can be rewritten as 
\[
\quad K_{3}=\frac{2T}{(2\pi T)^{3}}\frac{\left( -1\right) ^{3}}{2!}%
\mathop{\rm Re}%
\left( \Psi ^{(2)}(z)\right) ,\quad K_{5}=\frac{2T}{(2\pi T)^{5}}\frac{%
\left( -1\right) ^{5}}{4!}%
\mathop{\rm Re}%
\left( \Psi ^{(4)}(z)\right) \quad 
\]
and near the tricritical point, we have 
\begin{eqnarray*}
\alpha &=&-N(0)\cdot 
\mathop{\rm Re}%
\left( \Psi (\frac{1}{2}-i\frac{{\cal H}}{2\pi T})-\Psi (\frac{1}{2}-i\frac{%
{\cal H}_{0}}{2\pi T})\right) \\
&=&-N(0)\frac{({\cal H}-{\cal H}_{0})}{2\pi T}\cdot 
\mathop{\rm Im}%
\Psi ^{\prime }(\frac{1}{2}-i\frac{{\cal H}_{0}}{2\pi T})
\end{eqnarray*}
where ${\cal H}_{0}$ is the field corresponding to the second-order
transition into the uniform superconducting state, and it is given by\cite
{SST} 
\[
\ln \frac{T_{c}}{T}=%
\mathop{\rm Re}%
\left[ \Psi (\frac{1}{2}-i\frac{{\cal H}_{0}}{2\pi T})-\Psi (\frac{1}{2}%
)\right] . 
\]
Note also that in the expressions for all coefficients, except $\alpha $, in
the functional (\ref{GLF}) we may set ${\cal H}={\cal H}_{0}$.

\section{Minimization of the free energy}

Now, we proceed to study different solutions for the order parameter in one,
two and three dimensions, and see which solution minimizes the free energy.
We will compute the free energy when the order parameter is an exponential,
and a sine with one, two and three dimensional modulation.

\subsection{One-dimensional case}

If we choose the exponential order parameter 
\[
\psi (x)=\psi _{0}\;e^{iq\cdot x} 
\]

then the free energy reads 
\[
F_{\exp }=\left( \alpha +\beta q^{2}+\delta q^{4}\right) \cdot \left| \psi
_{0}\right| ^{2}+\left( \gamma +(\mu -2\eta )q^{2}\right) \cdot \left| \psi
_{0}\right| ^{4}. 
\]

Analysing the coefficient at $\left| \psi _{0}\right| ^{2}$ we see that the
field corresponding to the second-order transition depends on the wave
vector $q$, and that the actual field is the maximum one. Accordingly, we
get 
\begin{equation}
q_{\max }^{2}=-\frac{\beta }{2\delta },\qquad \alpha =\alpha _{0}=\frac{%
\beta ^{2}}{4\delta }  \label{condition}
\end{equation}

The free energy of such solution then reads 
\begin{equation}
F_{\exp }=\frac{-1}{2\pi N(0)}\frac{\left( \alpha -\alpha _{0}\right) ^{2}}{%
-K_{3}}  \label{OneDexp}
\end{equation}

Similarly, for the sine modulation $\psi (x)=\psi _{0}\;\sin (q.x)$, we find
the free energy 
\begin{equation}
F_{\sin }=\frac{-1}{\pi N(0)}\frac{\left( \alpha -\alpha _{0}\right) ^{2}}{%
-K_{3}}  \label{OneDcos}
\end{equation}

Note that since $K_{3}<0$ the denominators in these free energies are
positive in the region of existence of the FFLO state, which implies that
the transition is of second order. On the other hand, we see that the sine
modulation corresponds to the state with (twice) lower energy than the
exponential one in one dimension, and is therefore more favourable.

It is important to emphasize here that this sine solution for the order
parameter is just a limit of a more general solution on the transition line.
Indeed, we show below that as we drift away from the transition line, we
need to use an exact solution for the order parameter in one dimension\cite
{BP}.

Minimizing the free energy (\ref{GLF}) leads to following equation for the
order parameter (which is assumed to be a real function) 
\begin{equation}
\psi ^{(4)}+\widetilde{\alpha }\,\psi -\widetilde{\beta }\,\psi ^{\prime
\prime }-\widetilde{\mu }\,\left( \psi (\psi ^{\prime })^{2}+\psi ^{2}\psi
^{\prime \prime }\right) +2\,\widetilde{\gamma }\,\psi ^{3}+3\,\widetilde{%
\nu }\,\psi ^{5}=0  \label{L6}
\end{equation}
where we have redefined 
\[
\widetilde{\alpha }=\frac{\alpha }{\delta },\quad \widetilde{\beta }=\frac{%
\beta }{\delta },\quad \widetilde{\gamma }=\frac{\gamma }{\delta },\quad 
\widetilde{\nu }=\frac{\nu }{\delta },\quad \widetilde{\mu }=\frac{\mu
+2\eta }{\delta }. 
\]

Now using a similar approach to that of ref.\cite{BM}, we demonstrate that
the (Jacobi) elliptic-sine function 
\begin{equation}
\psi (x,k)=\Delta \,k\;sn(\frac{x}{\xi },k)  \label{soliton}
\end{equation}
is an adequate solution of eq.(\ref{L6}), $\Delta $ being its amplitude, $k$
the modulus of the elliptic sine and $\xi $ is some ''effective coherence
length'' that will be determined later on. Note that $sn(x,k)\rightarrow
\sin (x)$, as $k\rightarrow 0$, and this just happens on the transition
line, and that the amplitude of $\psi $ vanishes at the transition. Next,
owing to the Jacobi elliptic function \cite{GR}, $\psi (x,k)$ satisfies the
following equation\footnote{%
The prime here stands for the derivative with respect to the space
coordinate $x$.} 
\begin{equation}
\xi ^{2}(\psi ^{\prime })^{2}+(k^{2}+1)\psi ^{2}-\frac{\psi ^{4}}{\Delta ^{2}%
}=\Delta ^{2}k^{2}  \label{L7}
\end{equation}

Differentiating this equation with respect to $x$, we obtain 
\begin{equation}
\xi ^{2}\cdot \psi ^{\prime \prime }+(k^{2}+1)\cdot \psi -\frac{2}{\Delta
^{2}}\cdot \psi ^{3}=0  \label{L8}
\end{equation}
and differentiating the latter equation in turn twice and making use of it,
leads to the fourth-order equation, 
\begin{equation}
\psi ^{(4)}+\frac{(k^{2}+1)}{\xi ^{2}}\cdot \psi ^{\prime \prime }-\frac{12}{%
\xi ^{2}\Delta ^{2}}\cdot \left( \psi (\psi ^{\prime })^{2}+\psi ^{2}\psi
^{\prime \prime }\right) +\frac{12}{\xi ^{4}\Delta ^{4}}\cdot \psi ^{5}-%
\frac{6(k^{2}+1)}{\xi ^{4}\Delta ^{2}}\cdot \psi ^{3}=0  \label{L9}
\end{equation}

Multiplying (\ref{L7}) by $\psi $ and (\ref{L8}) by $\psi ^{2}$, and adding
the results leads to 
\begin{equation}
\frac{1}{\xi ^{2}\Delta ^{2}}\cdot \left( \psi (\psi ^{\prime })^{2}+\psi
^{2}\psi ^{\prime \prime }\right) -\frac{3}{\xi ^{4}\Delta ^{4}}\cdot \psi
^{5}+\frac{2(k^{2}+1)}{\xi ^{4}\Delta ^{2}}\cdot \psi ^{3}-\frac{k^{2}}{\xi
^{4}}\cdot \psi =0  \label{L10}
\end{equation}
and we also rewrite (\ref{L8}) as follows 
\begin{equation}
\frac{k^{2}+1}{\xi ^{2}}\cdot \psi ^{\prime \prime }+\frac{(k^{2}+1)^{2}}{%
\xi ^{4}}\cdot \psi -\frac{2(k^{2}+1)}{\xi ^{4}\Delta ^{2}}\cdot \psi ^{3}=0
\label{L11}
\end{equation}

Now, the point is to obtain an equation for $\psi $ that could be identified
with eq.(\ref{L6}). For this purpose, we add eq.(\ref{L9}) to the results of
multiplying respectively equations (\ref{L11}) and (\ref{L10}) by arbitrary
coefficients $A$ and $B$. Then by identifying the coefficients in the
resulting equation with those in (\ref{L6}), we obtain the following system
of $5$ equations for the $5$ parameters $\Delta ,\xi ,k,A$, and $B$: 
\begin{eqnarray}
\widetilde{\alpha } &=&\frac{A(k^{2}+1)^{2}-Bk^{2}}{\xi ^{4}},\quad 
\widetilde{\beta }=-\frac{(1+A)(k^{2}+1)}{\xi ^{2}}  \label{L12} \\
\widetilde{\gamma } &=&\frac{(B-A-3)(k^{2}+1)}{\xi ^{4}\Delta ^{2}}, 
\nonumber \\
\widetilde{\mu } &=&\frac{12-B}{\xi ^{2}\Delta ^{2}},\quad \widetilde{\nu }=%
\frac{4-B}{\xi ^{4}\Delta ^{4}}  \nonumber
\end{eqnarray}

On the other hand, in one dimension the coefficients $\widetilde{\alpha },%
\widetilde{\beta },\widetilde{\gamma },\widetilde{\mu }$ and $\widetilde{\nu 
}$ are obtained from eq.(\ref{GLF}), 
\begin{eqnarray}
\widetilde{\alpha } &=&\frac{16(K_{1}-K_{1}^{0})}{V_{F}^{4}K_{5}},\quad 
\widetilde{\beta }=-\frac{4K_{3}}{V_{F}^{2}K_{5}}  \label{L13} \\
\widetilde{\gamma } &=&-\frac{4K_{3}}{V_{F}^{4}K_{5}},  \nonumber \\
\widetilde{\mu } &=&\frac{10}{V_{F}^{2}},\quad \widetilde{\nu }=\frac{2}{%
V_{F}^{4}}  \nonumber
\end{eqnarray}

In addition, equating the ratio $\frac{\widetilde{\mu }^{2}}{\widetilde{\nu }%
}$ in (\ref{L12}) with that in (\ref{L13}) we infer that $B=2$ and $%
V_{F}^{2}=\xi ^{2}\Delta ^{2}$.

Moreover, as explained above, on the line of second-order transition the
solution must be $\sin (x)$, i.e. $k\rightarrow 0$, whereupon the
coefficients in (\ref{L12}) become 
\begin{eqnarray}
\widetilde{\alpha } &=&\frac{A}{\xi ^{4}},\quad \widetilde{\beta }=-\frac{1+A%
}{\xi ^{2}},\quad \widetilde{\gamma }=\frac{B-A-3}{\xi ^{4}\Delta ^{2}},
\label{L14} \\
\widetilde{\mu } &=&\frac{12-B}{\xi ^{2}\Delta ^{2}},\quad \widetilde{\nu }=%
\frac{4-B}{\xi ^{4}\Delta ^{4}}  \nonumber
\end{eqnarray}

Then, using the condition for the second-order transition, 
\begin{equation}
\widetilde{\alpha }=\frac{\widetilde{\beta }^{2}}{4}  \label{L15}
\end{equation}
we find that $A=1$.

At this point we wish to stress that in two and three dimensions one can
readily check that the elliptic sine is no longer a good solution that
transforms into sine on the transition line. Indeed, on this line $A=1,$ so
that the identity $\widetilde{\gamma }=\frac{\widetilde{\beta }}{V_{F}^{2}}$
infered from eq.(\ref{L14}) is compatible with eq.(\ref{L13}) that is valid
in one dimension. However, this is not true in other dimensions since in
this identity the right hand side depends on the dimension of space whereas
the left hand side doesn't, see (\ref{GLF}).

Away from the transition line defined by eq.(\ref{L15}), the parameter $A$
obtained from (\ref{L12}) can be written in terms of the ''external field''
parameter $h=\frac{\widetilde{\alpha }}{\widetilde{\beta }^{2}}$ as follows 
\begin{equation}
A(k,h)=-1+\frac{1}{2h}-\frac{1}{2h}\sqrt{1-4h\;\left( 1+\frac{2k^{2}}{%
(k^{2}+1)^{2}}\right) }  \label{A}
\end{equation}

Of course, $A(k,h)\rightarrow 1$ as $k\rightarrow 0$, since then $%
h\rightarrow \frac{1}{4}$ due to eq.(\ref{L15}).

Note that in this case we have in fact a family of solutions parametrized by 
$k$. Indeed, using eqs.(\ref{L12}), (\ref{L13}) we can express the
parameters $\Delta $ and $\xi $ in terms of the modulus $k$, bearing in mind
that $B=2$ and $V_{F}^{2}=\xi ^{2}\Delta ^{2}$, see above.

Therefore, in one dimension we have found a solution to the system of
equations (\ref{L12}) for all parameters entering in the order parameter $%
\psi (x,k)$ described by the general Ginzburg-Landau functional (\ref{GLF}).
Now, let us compute the free energy of this non-uniform order parameter and
compare it with that of the uniform state. We shall also study the
transition between these two states. For this purpose, we insert the
solution (\ref{soliton}) into the Ginzburg-Landau functional (\ref{GLF}) and
integrate over a period of the Jacobi elliptic sine, namely $4K(k)$, $K(k)$
being the complete elliptic integral of first kind\cite{GR}, thus we obtain
the following expression 
\begin{equation}
\frac{F_{NU}(k,h)}{C}=\frac{-k^{4}}{\left[ (k^{2}+1)(A+1)\right] ^{3}}\left[
10\;I_{2}-(k^{2}+1)(11+A)\;I_{4}+14k^{2}\;I_{6}\right]  \label{FNU}
\end{equation}
where we have defined 
\[
I_{n}(k)=\frac{1}{K(k)}%
\displaystyle \int %
\limits_{0}^{\frac{\pi }{2}}\frac{dx\sin ^{n}x\,}{\sqrt{1-k^{2}\sin ^{2}x}} 
\]
or in terms of $K(k)$ and the complete elliptic integral of second kind $%
E(k) $ , 
\begin{eqnarray*}
I_{2} &=&\frac{1}{k^{2}}\left( 1-\frac{E(k)}{K(k)}\right) ,\quad I_{4}=\frac{%
2+k^{2}}{3k^{4}}-\frac{2(1+k^{2})}{3k^{4}}\frac{E(k)}{K(k)} \\
I_{6} &=&\frac{4(k^{2}+1)}{5k^{2}}\;I_{4}-\frac{3}{5k^{4}}\left( 1-\frac{E(k)%
}{K(k)}\right)
\end{eqnarray*}
The coefficient $C=\delta \;V_{F}^{2}\;(-\widetilde{\beta })^{3}$ is
positive since $\widetilde{\beta }<0$.

Note that the most advantageous feature of the free energy (\ref{FNU}) is
that it depends only on the parameter $k$ once the external field parameter $%
h$ is fixed. Then, upon minimizing this free energy we determine the modulus 
$k$ and thereby we obtain the parameters $\xi $ and $\Delta $ from eq.(\ref
{L12}), 
\[
\xi ^{2}=-\frac{(1+A(k,h))(k^{2}+1)}{\widetilde{\beta }},\quad \Delta =\frac{%
V_{F}}{\xi } 
\]

On the other hand, the free energy for the uniform order parameter reduces
to 
\[
F_{U}=\delta \cdot \left( \widetilde{\alpha }\,\psi ^{2}+\widetilde{\gamma }%
\,\psi ^{4}+\,\widetilde{\nu }\,\psi ^{6}\right) 
\]
where the coefficient $\widetilde{\gamma }$ is negative, and hence the
transition is of first order here.

Using the same notation as above, the free energy $F_{U}$ becomes 
\begin{equation}
\frac{F_{U}(h)}{C}=\frac{-1}{18}\left[ (1+\sqrt{1-6h})(\frac{1}{3}%
-2h)-h\right]  \label{FU}
\end{equation}
which depends on the sole parameter $h$.

It is seen in figure 1 that the free energies (\ref{FNU}) and (\ref{FU})
converge when the external field $h$ becomes equal to the critical value $%
hc\simeq 0.0925925$. The latter marks the transition from the non-uniform to
uniform state. Next, we show in figure 2 the change of the form and period
of the order parameter from sine to soliton lattice as we lower the external
field. When $h\leq h_{c}$, the period of the order parameter is infinite and
the superconducting state becomes uniform, see figure 3.

Before considering the two and three dimensional cases, let us first compute
the jump in the specific heat at the second-order transition in one
dimension. In this case the free energy for the sine solution is given in
eq.(\ref{OneDcos}), and the corresponding jump in the specific heat reads 
\begin{equation}
\Delta C=4N(0)T_{c}\times \frac{\left( 
\mathop{\rm Im}%
\Psi ^{\prime }(\frac{1}{2}-i\frac{{\cal H}_{0}}{2\pi T})\cdot \frac{d{\cal H%
}_{0}}{dT}\right) ^{2}}{%
\mathop{\rm Re}%
\Psi ^{(2)}(\frac{1}{2}-i\frac{{\cal H}_{0}}{2\pi T})}  \label{L16}
\end{equation}

The characteristic feature of such behaviour is the divergence of the jump
at the tricritical point.

\subsection{Two-dimensional case}

\subsubsection{Exponential}

As above, we start by considering the exponential order parameter 
\[
\psi (r)=\psi _{0}\;e^{iq\cdot r} 
\]
for which the free energy reads 
\[
F=\left( \alpha +\beta q^{2}+\delta q^{4}\right) \cdot \left| \psi
_{0}\right| ^{2}+\left( \gamma +(\mu -2\eta )q^{2}\right) \cdot \left| \psi
_{0}\right| ^{4}, 
\]
and upon minimizing over $q^{2}$ and $\left| \psi _{0}\right| ^{2}$, this
becomes 
\begin{equation}
F_{\exp }=\frac{1}{\pi N(0)}\frac{\left( \alpha -\alpha _{0}\right) ^{2}}{%
K_{3}}.  \label{TwoDexp}
\end{equation}

For one-component sine 
\[
\psi (r)=\psi _{0}\;\sin (q.x) 
\]
we obtain 
\begin{equation}
F_{\sin _{x}}=\frac{6}{\pi N(0)}\frac{\left( \alpha -\alpha _{0}\right) ^{2}%
}{K_{3}}  \label{2Dcosx}
\end{equation}
and finally for the two-component sine 
\[
\psi (r)=\psi _{0}\;\left( \sin (q.x)+\sin (q.y)\right) 
\]
we get 
\begin{equation}
F_{\sin _{xy}}=\frac{4}{\pi N(0)}\frac{\left( \alpha -\alpha _{0}\right) ^{2}%
}{K_{3}}  \label{2Dcos_xy}
\end{equation}
Therefore, since $K_{3}<0$, we infer that 
\[
F_{\sin _{x}}<F_{\sin _{xy}}<F_{\exp } 
\]

This means that the phase with one-component sine modulation is the most
stable in two dimensions. Here again we find that the coefficients of the
quartic term in the free energy are positive, which implies that the
transition into the FFLO state is of second-order{\bf . }Our result can
justify the choice made in ref.\cite{Burkh} of the structure depending only
on one coordinate for the numerical analysis of the FFLO phase in two
dimensions.

Finally, as in one dimension we obtain the jump in the specific heat 
\begin{equation}
\Delta C=24N(0)T_{c}\times \frac{\left( 
\mathop{\rm Im}%
\Psi ^{\prime }(\frac{1}{2}-i\frac{{\cal H}_{0}}{2\pi T})\cdot \frac{d{\cal H%
}_{0}}{dT}\right) ^{2}}{%
\mathop{\rm Re}%
\Psi ^{(2)}(\frac{1}{2}-i\frac{{\cal H}_{0}}{2\pi T})}\quad  \nonumber
\end{equation}
which diverges at the tricritical point.

\subsection{Three-dimensional case}

The free energy of the exponential order parameter $\psi (r)=\psi
_{0}\;e^{iq\cdot r},$ reads 
\begin{equation}
F=\left( \alpha +\beta q^{2}+\delta q^{4}\right) \cdot \left| \psi
_{0}\right| ^{2}+\left( \gamma +(\mu -2\eta )q^{2}\right) \cdot \left| \psi
_{0}\right| ^{4}.  \label{3DExp}
\end{equation}

Then using the coefficients in eq.(\ref{GLF}) we see that the coefficient of
the quartic term in (\ref{3DExp}) is positive at $q^{2}=q_{0}^{2}=-\frac{%
\beta }{2\delta }$, thus indicating that the transition is of second order.

On the contrary, for the $1,2$, and $3$ component sine we find that this
coefficient is negative and consequently the transition is of first order.
But, in order to compute the critical temperature it is necessary to add the
sixth-order term $\upsilon \psi ^{6}$. The transition occurs into a
structure with well-developped harmonics and to find the corresponding
critical temperature and the structure of the non-uniform state numerical
calculations are needed.

The functional (\ref{GLF}) provides a good basis for such calculations.
Apparently the $3D$-lattice state is more stable. Really, one could also
study the change in the order of the transition as we drift toward the $T=0$
point. In fact, it was argued in \cite{Malaspinas} that as $T\rightarrow 0$
the transition into the sine state becomes of second order, whilst the
results of ref.\cite{LO} showed that at this point the transition into the $%
3D$-lattice state remains of first order, and thus corresponds to a higher
critical field and lower energy as compared with the sine state with
one-dimensional modulation.

We may speculate that the $3D$-lattice is also more favourable near the
tricritical point and along the whole transition line.

\section{Conclusion}

We have constructed a generalized Ginzburg-Landau functional for the FFLO
superconductors near the tricritical point in the $(T,H)$-phase diagram, in $%
1,2$, and $3$ dimensions. Using this functional we have shown that the state
with exponential modulation of the order parameter ($\psi \sim e^{iqx}$) is
always unfavourable and that the transition from the normal to the FFLO
state is of second order in $1$ and $2$ dimensions, where the order
parameter has a one-component sine modulation near the transition line. In
one dimension we have shown that upon lowering the external field, the
order-parameter modulation changes from $\sin (x)$ on the transition line
into elliptic sine $sn(x),$ and finally transfers to a uniform state. The
proposed generalized Ginzburg-Landau functional for $1D$ superconductors can
be directly applied to the description of the spin-Peierls systems in
magnetic field near the corresponding tricritical point\cite{RB}.

In $3$ dimensions we have shown that the transition is of first order, and
contrary to $1$ and $2$ dimensions, one has to take into account higher
harmonics of the order parameter to calculate the critical temperature.

\section{Acknowledgement}

We are thankful to J.-P. Brison and J. Flouquet for helpful discussions.

\newpage 

{\Large Figure captions}

\begin{itemize}
\item  Figure 1: Plot of free energy as a function of the ''external field'' 
$h=\frac{\widetilde{\alpha }}{\widetilde{\beta }^{2}},C=\delta
\;V_{F}^{2}\;(-\widetilde{\beta })^{3}$. The solid line represents the free
energy of the uniform state. The line in crosses represents the free energy
of the non-uniform soliton-lattice state. These two curves meet at $%
h_{c}=0.0925925$.
\end{itemize}

\vspace{1cm}

\begin{itemize}
\item  Figure 2 : Spatial dependence of the superconducting order parameter
in the non-uniform $1D$ phase for different fields $h=\frac{\widetilde{%
\alpha }}{\widetilde{\beta }^{2}}$. The dotted line indicates the solution
at $h=0.165$, and the solid line represents the solution when $h=0.0927$,
i.e. close to the critical value $h_{c}\simeq 0.0925925$. The dimensionless
space coordinate $\widetilde{x}$ is defined by $\widetilde{x}=x\cdot \left( 
\sqrt{-\widetilde{\gamma }}V_{F}\right) ,\quad $and $\widetilde{\psi }(%
\widetilde{x},k)=\frac{\psi (\widetilde{x},k)}{\sqrt{(1+A(k,h))(k^{2}+1)}}.$
\end{itemize}

\vspace{1cm}

\begin{itemize}
\item  Figure 3 : Variation of the period $L$ of the non-uniform phase as a
function of the external field $h$. The period diverges as the external
field approaches the critical value $h_{c}$. The dimensionless period $%
\widetilde{L}$ plotted here is defined as $\widetilde{L}=L\cdot \left( \sqrt{%
-\widetilde{\gamma }}V_{F}\right) $, and for the soliton-lattice phase this
is $\widetilde{L}=4K(k)\cdot \sqrt{(1+A(k,h))(k^{2}+1)}$, see text.
\end{itemize}

\end{document}